\documentstyle[sprocl,epsfig]{article}

\def\Journal#1#2#3#4{{#1} {\bf #2}, #3 (#4)}

\def\NIMA{{\em Nucl. Instrum. Methods} A}

\def\be{\begin{equation}}
\def\ee{\end{equation}}
\def\bea{\begin{eqnarray}}
\def\eea{\end{eqnarray}}

\begin{document}


\title{NEW DEVELOPMENTS OF PHOTODETECTORS FOR THE LAKE BAIKAL
NEUTRINO EXPERIMENT}

\author{B. K. LUBSANDORZHIEV}

\address{Institute for Nuclear Research of Russian Academy of Sciences, pr-t 60-letiya Oktyabrya 7A,
\\ Moscow 117312, Russia\\E-mail: lubsand@pcbai10.inr.ruhep.ru}




\maketitle\abstracts{ New developments of photodetectors for the lake
Baikal neutrino experiment are described. Some test results of
photodetectors at the lake Baikal are presented.}

\section{Introduction}
The lake Baikal Neutrino Experiment has history of more than 20 years,
starting from small short experiments with a few PMTs in the early 80s to
the present large scale longterm operating neutrino telescope NT-200\cite{APP},
which has been put into operation on April 6th 1998. The telescope's
effective area for muons is 2000-10000 depending on a muon energy.
The rate of events due to atmospheric neutrinos is about 1
per two days.

\section{Neutrino telescope NT-200}

The lake Baikal Neutrino Telescope NT-200 is located in the southern part of the great
Siberian lake Baikal at 3.6km from the shore and at the depth of 1km.
The schematic view of NT-200 and optical modules attached at a string are presented in fig.1.
The telescope consists of 192
optical modules at 8 vertical strings arranged at an umbrella like frame.
Optical modules are grouped in pairs and swithched in coincidence with 15ns
time window. So two optical modules in one pair define one \mbox{\it{optical channel}}
resulting in rather low \mbox{\it{optical channel}} background counting rate of 100--300Hz.
Two pairs of optical modules form so called \mbox{\it{svyazka}}.
The detector electronics system is hierarchical: from the lowest level
to the highest one - optical module electronics, \mbox{\it{svyazka}} electronics
module, string and detector electronics modules. In the latter detector
trigger signals are formed and all information from string electronics
modules are received and sent to the shore station. Three underwater
electrical cables connect the detector with the shore station.
The detector is operated from the shore station.

\begin{figure}[t]
\centering
\mbox{\psfig{figure=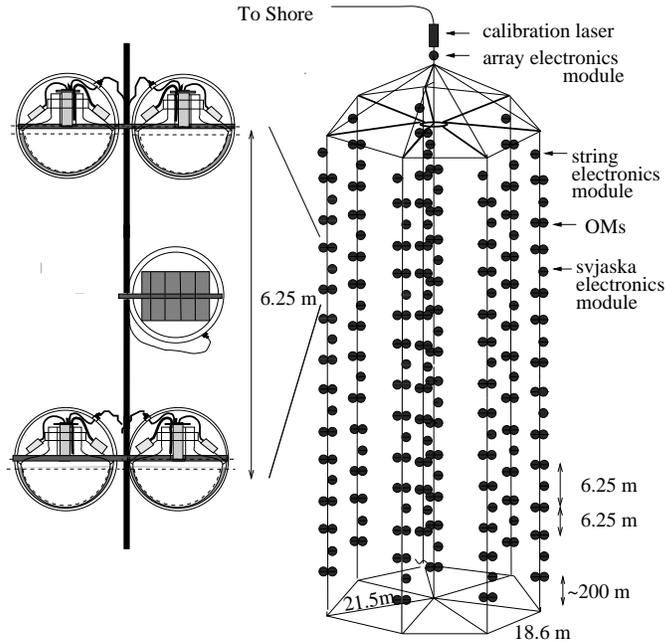,height=3.5in,angle=-90}}
\caption{The lake Baikal neutrino telescope NT-200}
\end{figure}

\section{Quasar-370 phototube}

\begin{figure}[t]
\centering
\mbox{\psfig{figure=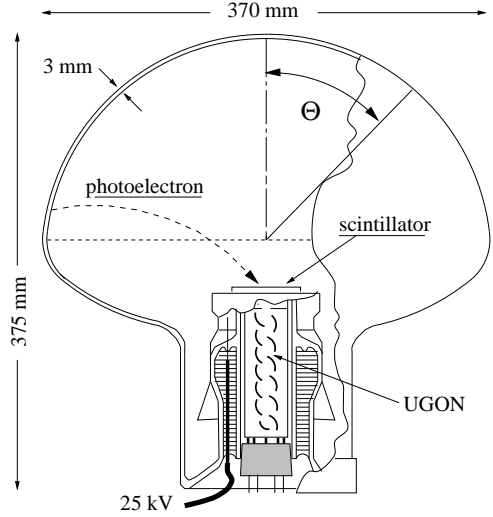,height=2.5in,angle=-90}}
\caption{Quasar-370 phototube}
\end{figure}

The Quasar-370 phototube \cite{Pil,Aa,NIMBAGD,NIMLBK}
is a hybrid phototube which
consists of an
electro-optical preamplifier with large area hemispherical photocathode
and small conventional type PMT, see fig.2. Photoelectrons from a large 37cm
diameter
hemispherical photocathode with $2\pi$ acceptance  are accelerated by 25 kV
to a fast, high gain luminescent screen ($YSO$ scintillator is usually used).
The light flashes in luminescent
screen induced by photoelectrons are read out by small PMT
with 3cm diameter photocathode.
The latter has been
developed especially for this kind of application by INR and MELZ factory
\cite{UGON}.
As a result one
photoelectron from the hemispherical photocathode yields typically 25
photoelectrons in the small PMT.This high gain first stage results in an
excellent single photoelectron resolution. Due to the fast acceleration of
primary photoelectrons by 25 kV high voltage and mushroom shaped glass
envelope the time jitter can be kept rather low. Last but not the least the tube
is almost insensitive to the Earth's magnetic field.
Averaged over more than 200 tubes the mean values for single photoelectron
resolution and time resolution is 70\% and 2ns respectively.
Quasar-370 phototube is characterized by conspicuously low level of
afterpulses in comparison with conventional PMTs. The level of afterpulses
in Quasar-370 tube is substantially less than 1\%. Another advantage of
Quasar-370 phototube in contrary to conventional PMTs is the lack of
prepulses due to the fact that the first stage of the phototube is
optically separated from tube's photocathode.
It was shown \cite{Pil,Aa,NIMLBK} that Quasar-370 parameters depend strongly
on characteristics of a scintillator in a luminescent screen.
So recently we have developed a number of modifications of Quasar-370 tube
with new scintillators. The most promising results we have got with $ScBO_3$,
$YAP$ and $LSO$. Unfortunately the latter one has one substantial drawback for
using in Quasar-370 tube. It has rather high value of $Z_{eff}$. It is important
to note here we need scintillators with $Z_{eff}$ as low as possible to suppress
effectively \mbox{\it{late}} pulses \cite{NIMLATE} due to photoelectron backscattering
effect. With these new scintillators we have reached $~1$ns (FWHM) time jitter
and $~40$\% single photoelectron resolution. It is noteworthy to mention here
about very much intriguing scintillator\cite{Lucky} - $ZnO:Ga$ with less than 1ns decay
time and light yield of 40\% of $NaI:Tl$. At present we  try
to work with this scintillator and to reproduce results of \cite{Lucky}.
The success of this work would undoubtedly be a kind of a breakthrough in designing of
very fast and effective hybrid phototubes.

\section{Two-channel optical module}

The pairwise ideology pursued in NT-200 neutrino telescope has
many advantages. The ideology allows to suppress effectively an individual
optical module background counting rate due to water luminescence and a
phototube's dark current, to eliminate phenomena deteriorating phototube's
time resolution namely \mbox{\it{prepulses}}, \mbox{\it{late}} pulses and
\mbox{\it{afterpulses}}. Moreover such
an approach facilitates very much designing of trigger system, data acquisition
system etc. Unfortunately there is just one but essential shortcoming.
Namely it's too expensive to have two optical modules for one
\mbox{\it{optical channel}}.
To overcome this problem we have developed two-channel optical module based
on two-channel version of Quasar-370 phototube\cite{DUR}.

\begin{figure}[t]
\centering
\mbox{\psfig{figure=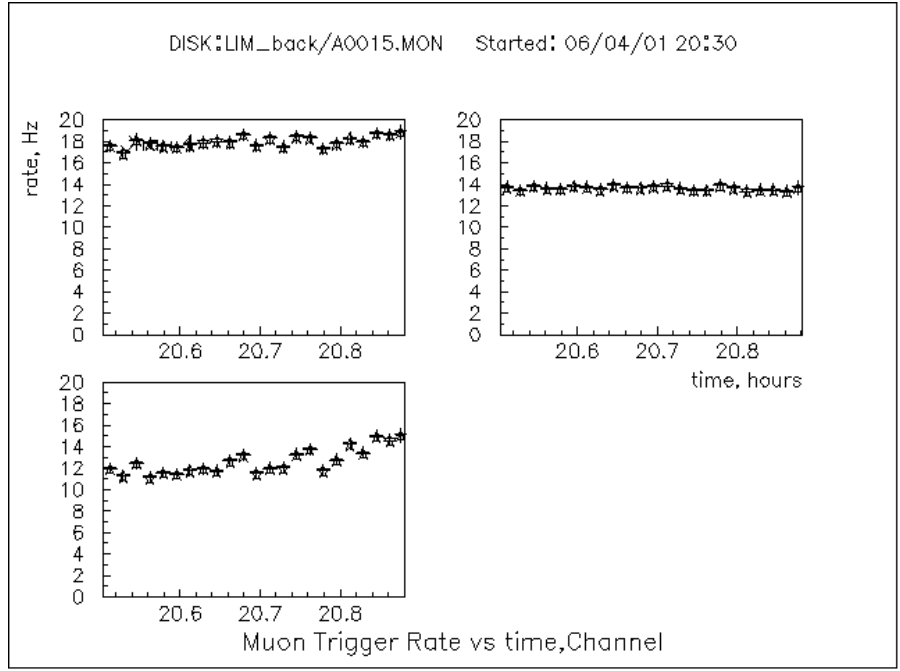,height=3.5in}}
\caption{Muon Trigger rate versus time. Left up and left bottom -
for conventional Baikal \mbox{\it{optical channel}}. Right up - for new
\mbox{\it{optical channel}} based on new two-channel optical module.}
\end{figure}

Two-channel Quasar-370 phototube uses the same electro-optical preamplifier
and new two-anode small PMT with mesh dynode system instead of conventional
PMT. This new two-anode
small PMT has following characteristics: time jitter -- $\sigma=340$ps;
Peak-to valley ratio of single photoelectron sharge distribution --  1.5;
cross-talks between channels --  1\%.
Basing on two-channel Quasar-370 phototube we have developed two-channel
optical module which incorporates besides Quasar-370 phototube two fast transimpedance
preamplifiers for each anode signals and uses the same glass pressure vessel
and the same penetrators as conventional Baikal optical module.
In the course of the last expedition at the lake Baikal we have tested
one pilot sample of two-channel optical module in frame of a new experimental
string designed to test technological innovations for future neutrino telescopes
at the lake. In the test measurements at the lake we used ordinary but slightly
modified NT-200 front-end electronics. Output signals of each channel are
switched in coincidence just in the same way as in NT-200. It results in the
same \mbox{\it{local trigger}} counting rate of 100-300Hz. Fig.3 presents
the time dependence of muon trigger rate for two conventional Baikal
\mbox{\it{optical channels}} and new \mbox{\it{optical channel}} based on
two-channel optical module. One can see that the new \mbox{\it{optical channel}}
has comparable sensitivity to muons with ordinary \mbox{\it{optical channels}}
of NT-200. It opens new possibilities for NT-200 further extension plans.

\section{Other developments}

A new version of Quasar-370 phototube (Quasar-370D) with semiconductor diode as a
photoelectron multiplying element instead of a system of 
luminescent screen and small PMT has been developed. We have manufactured two
pilot samples which are under studies now.
Modifications of Quasar-370 phototubes (Quasar-370G)\cite{PTE} which are able to
withstand high current due to night sky background operate successfully  in
the wide angle atmospheric air Cherenkov detectors TUNKA\cite{Tun} and SMECA\cite{Ya}.
Quasar-370L is low background version of Quasar-370 phototube. It has $U^{238}$ and $Th^{232}$ content of about $10^{-8}$g/g. Quasar-370L is aimed at using in low background experiments. 
Another development is concerned with phototubes with high quantum efficiency
(more than 40\% in the range 450-550nm) photocathodes and it's results will be
reported elsewhere.

\section*{Conclusion}
The lake Baikal Neutrino Experiment successful operation proves high performance and high
reliability of a number of photodetectors developed for this kind of
application. New developments are focused at future neutrino experiments at the lake
Baikal and other experiments in high energy physics, cosmic ray and astroparticle physics.

\section*{Acknowledgments}
This work was supported by the Russian Ministry of Research
(contract 102-11(00)-p), the German Ministry of Education and Research and
the Russian Fund of Basic
Research (grants 99-02-1837a,01-02-31013 and 00-15-96794)
and by the Russian Federal Program "Integration"(project 346).
The author is indebted very much to his colleagues from KATOD and MELZ laboratoties and 
Baikal Collaboration.
The author would like to thank particularly Dr. V.Ch.Lubsandorzhieva for many invaluable advices and help in preparation of this paper.


\section*{References}
\begin{thebibliography}{99}
\bibitem{APP} I.A.Belolaptikov {\it et al},
{\it Astropart. Phys.} {\bf 7} {263} {1997}
\bibitem{Pil} L.B.Bezrukov {\it et al} Proc. 3rd NESTOR Workshop, Pylos 1993.
Univ.Athens. {132} {1994}
\bibitem{Aa} R.I.Bagduev {\it et al} Proc. Int.Conf. "Trends in Astroparticle
Physics", Teubner, Stutgart, {132} {1994}
\bibitem{NIMBAGD} R.I.Bagduev {\it et al}, \Journal{\NIMA}{420}{138}{1999}
\bibitem{NIMLBK} B.K.Lubsandorzhiev, \Journal{\NIMA}{442}{368}{2000}
\bibitem{UGON} L.B.Bezrukov {\it et al}, {\it Instrum. and Experim. Techn.}
{\bf 1} {104} {2000}
\bibitem{NIMLATE} B.K.Lubsandorzhiev {\it et al}, \Journal{\NIMA}{442}{452}{2000}
\bibitem{Lucky} PHILIPS Photomultiplier tubes, {6-35} {1994}

\bibitem{DUR} B.K.Lubsandorzhiev {\it et al}, Proc. of the 25th ICRC.{\bf 7} {269} {1997}

\bibitem{PTE} B.K.Lubsandorzhiev {\it et al}, {\it Instrum. and Experim. Techn.}
{\bf 3} {104} {2001}

\bibitem{Tun} N.Budnev {\it et al}, Proc. of the 25th ICRC.{\bf 2} {581} {2001}

\bibitem{Ya} V.A.Balkanov {\it et al}, {\it Yadernaya Fizika} {\bf 63} {1027} {2000}
\end{thebibliography}
\end{document}